\documentstyle[preprint,eqsecnum,aps,epsf]{revtex}

\catcode`\@=11
\def\@citex[#1]#2{%
\if@filesw \immediate \write \@auxout {\string \citation {#2}}\fi
\@tempcntb\m@ne \let\@h@ld\relax \def\@citea{}%
\@cite{%
  \@for \@citeb:=#2\do {%
    \@ifundefined {b@\@citeb}%
      {\@h@ld\@citea\@tempcntb\m@ne{\bf ?}%
      \@warning {Citation `\@citeb ' on page \thepage \space
undefined}}%
      {\@tempcnta\@tempcntb \advance\@tempcnta\@ne%
      \@tempcntb\number\csname b@\@citeb \endcsname \relax%
      \ifnum\@tempcnta=\@tempcntb %   Number follows previous--hold on to it
	\ifx\@h@ld\relax%
	  \edef \@h@ld{\@citea\csname b@\@citeb\endcsname}%
	\else%
	  \edef\@h@ld{\ifmmode{-}\else--\fi\csname
b@\@citeb\endcsname}%
	\fi%
      \else%   %  non-successor--dump what's held and do this one
	\@h@ld\@citea\csname b@\@citeb \endcsname%
	\let\@h@ld\relax%
      \fi}%
    \def\@citea{,\penalty\@highpenalty\,}%
  }\@h@ld
}{#1}}

\def\@citeb#1#2{{[#1]\if@tempswa , #2\fi}}
\def\@citeu#1#2{{$^{#1}$\if@tempswa , #2\fi }}
\def\@citep#1#2{{#1\if@tempswa , #2\fi}}
\def\bcites{         % cite with []'s
	\catcode`\@=11
	\let\@cite=\@citeb
	\catcode`\@=12
}
\def\upcites{         % cite with exponents
	\catcode`\@=11
	\let\@cite=\@citeu
	\catcode`\@=12
}
\def\plaincites{      % cite without brackets
	\catcode`\@=11
	\let\@cite=\@citep
	\catcode`\@=12
}

\begin{document}
\draft

\title{Flux-Tube Ring and Glueball Properties \\
in the Dual Ginzburg-Landau Theory}

\author{Y. Koma\footnote{E-mail address: koma@rcnp.osaka-u.ac.jp},  
H. Suganuma and H. Toki}

\address{Research Center for Nuclear Physics (RCNP), Osaka University\\
Mihogaoka 10-1, Ibaraki, Osaka 567-0047, Japan}

\date{\today} 
\maketitle

\begin{abstract}
An intuitive approach to the glueball using the
flux-tube ring solution in the dual Ginzburg-Landau theory 
is presented.
The description of the flux-tube ring as the relativistic 
closed string with the effective string tension enables us to 
write the hamiltonian of the flux-tube ring using the Nambu-Goto action.
Analyzing the Schr$\ddot{\rm {o}}$dinger equation, we discuss the mass 
spectrum and the wave function of the glueball.
The lowest glueball state is found to have the mass 
$M_G \sim 1.6 \;{\rm GeV}$ and the size $R_G \sim 0.5 \;{\rm fm}$. \\
\end{abstract}


\baselineskip = 0.75cm
{\small
%%% SECTION 1 %%%%%%%%%%%%%%%%%%%%%%%%%%%%%%%%%%%%%%%%%%%%%%%
\section{Introduction}
\label{sec:intro}

\par
The existence of glueball states is naively expected in terms of 
the gluon self-coupling in QCD\cite{Creutz}.
Recent progress of lattice QCD simulations predicts the masses of glueballs 
$M(0^{++})$ = 1.50 $\sim$ 1.75 GeV, $M(2^{++})$ = 2.15 $\sim$ 2.45 GeV
\cite{Bali,Chen,Sexton,Teper,Morningstar}.
Experimentally, there are some candidates;
$f_0(1500)$ and $f_0(1710)$ for the scalar glueball,
$f_J(2220)$ ($J$=2 or 4), $f_2(2300)$ and $f_2(2340)$ for the 
tensor glueball\cite{PDG}.
However, the abundance of $q$-$\bar{q}$ meson states in the 1 $\sim$ 3 GeV
region and the possibility of the quarkonium-glueball mixing states still
make it difficult to identify the glueball states\cite{Tornqvist}.
To date, no glueball state has been firmly discovered yet.
More studies for the glueballs from many directions are necessary to 
specify the glueball states.

\par
In this paper, we present an analytic and very intuitive 
approach to the glueball using the dual Ginzburg-Landau (DGL) theory,
an effective theory of the nonperturbative QCD.
The DGL theory is constructed from QCD by performing
the 't Hooft abelian projection\cite{tHooft2} with 
the two hypotheses, abelian dominance and monopole condensation\cite{Ezawa}.
Abelian projection reduces QCD into the ${\rm U(1)^2}$ abelian 
gauge theory including monopoles, and
recent studies of the lattice QCD in the maximally abelian (MA) 
gauge give numerical evidences of QCD-monopole 
condensation\cite{Kronfeld,Atanaka} and
abelian dominance\cite{Brandstater,Amemiya} for the 
nonperturbative phenomena such as 
confinement\cite{Hioki,Stack},
chiral symmetry breaking\cite{Miyamura,Woloshyn}. 
In the DGL theory, the QCD-vacuum is described as the dual version of the 
superconductor and the color confinement 
is realized by the formation of the color-electric flux-tube through the dual 
Meissner effect\cite{Suzuki,Suganuma,Ichie1,Monden}.
The flux-tube has a constant energy per unit length, the string 
tension, which characterizes the strength of the color confinement
as the slope of the linear potential between the color charges.
The flux-tube solution in the DGL theory appears as the topological 
excitation as the relevant collective mode in the QCD-vacuum, and this
provides intuitive pictures of the hadrons in terms of the string-like 
structure of the color-electric flux.
While the hadrons including the valence quarks
correspond to an open flux-tube excitation with terminals, 
the glueball can be regarded as the flux-tube without end, 
 the ``flux-tube ring'' excitation\cite{Koma}, in the flux-tube 
picture\cite{Isgur}.
This simple picture is expected to provide the further 
understanding of the glueball.

\par
In this paper, we consider the simplest ring solution that the ends 
of the flux-tube meet each other to form a circle. We study then the 
profiles of the color-electric field and the monopole field with the DGL 
theory.
We calculate also the ``effective string tension'', which is
the string tension of the flux-tube forming
a ring as a function of the ring radius $R$.
The flux-tube ring solution in the DGL theory itself is
unstable and prefers to shrink, since it does not 
contain any kinetic term for the ring motion. 
From the quantum mechanical point of view, such a collapse is to
be forbidden by the uncertainty principle.
Let us imagine the hydrogen atom, where the stable ground state
is determined by the energy balance between 
the kinetic term of the electron $p^2/2 m_e$ 
and the Coulomb potential term $-e^2/r$
with the uncertainty relation $p \cdot r \geq 1$.
Similarly, it would be necessary to introduce the kinetic term of the ring,
and take the quantum effect into account.
However, it is a difficult problem since we have no guiding 
principle to determine the kinetic term of the flux-tube.
Therefore, in this paper, we introduce the kinetic term based on the 
string-like description of the flux-tube by using the Nambu-Goto (NG) 
action in the string theory\cite{Green} and
we use the principle that the string action is proportional to the 
world surface swept over the string motion.
In this scheme, the flux-tube ring is regarded as the relativistic 
closed string with the effective string tension, which is calculated
based on the DGL theory.

\par
In section \ref{sec:DGLT}, we investigate the single flux-tube solution 
in the DGL theory and consider the essence of the flux-tube.
The DGL parameters are determined so as to reproduce the
string tension $\sigma \simeq 1\;{\rm GeV/fm}$ 
extracted from the Regge slope of the hadrons\cite{Regge}.

\par
In section \ref{sec:ring}, we study the flux-tube ring solution as 
the glueball excitation. 
We investigate the profiles and the effective string tension
of the flux-tube ring, and then, we combine the DGL theory with 
the string theory in order to introduce the kinetic term and write the
hamiltonian of the flux-tube ring.
Finally, we estimate the mass and the size of the glueball by solving the
Schr$\ddot{\rm o}$dinger equation.

\par
Section \ref{sec:summary} is devoted to the summary of the present
study and the discussion.

%%% SECTION 2 %%%%%%%%%%%%%%%%%%%%%%%%%%%%%%%%%%%%%%%%%%%%%%%
\section{Single Flux-Tube Solution in the DGL theory}
\label{sec:DGLT}

In this section, we consider the topological solution related to 
the $q$-$\bar{q}$ system in the dual Ginzburg-Landau (DGL) theory
within the quenched level.
The system is described by the DGL lagrangian\cite{Suzuki,Suganuma},
\begin{equation}
{\cal L}_{\rm DGL} =
-\frac{1}{4} \left (\partial_{\mu}\vec{B}_{\nu}-\partial
_{\nu}\vec{B}_{\mu} 
-\frac{1}{n \cdot \partial}\varepsilon_{\mu \nu \alpha \beta}
n^{\alpha} \vec{j}^{\beta} \right )^2
  +\sum_{\alpha=1}^3 \left [ \left | \left (\partial_{\mu}+ig\vec{\epsilon}
_{\alpha}{\cdot}\vec{B}_{\mu} \right )\chi_{\alpha} \right |^2
-\lambda \left ( \left |\chi_{\alpha} \right |^2-v^2 \right )^2 \right ],
\label{eqn:DGL}
\end{equation}
where $\vec{B}_{\mu}$ and $\chi_{\alpha}$ denote the dual gauge field with 
two components $(B_{\mu}^3, B_{\mu}^8)$ and the complex scalar monopole 
field, respectively. Here, $\vec{\epsilon}_a$ is the root 
vector of SU(3) algebra,
$\vec{\epsilon}_1=\left (-1/2,\sqrt{3}/2 \right ), \vec{\epsilon}_2=\left
(-1/2,-\sqrt{3}/2 \right )
, \vec{\epsilon}_3=\left (1,0 \right )$, and $n^{\mu}$ denotes 
an arbitrary constant 4-vector, which corresponds
to the direction of the Dirac string.
At the quenched level, the color sources are given as the $c$-number 
current, and the heavy $q$-$\bar{q}$ system provides
\begin{equation}
{\vec{j}^{\mu}}_{\alpha}(x)
\equiv \vec{Q}_{\alpha} g^{\mu 0} 
\left[ 
\delta^3 \left (\mbox{\boldmath $x$} 
- \mbox{\boldmath $a$} \right )
- \delta^3  \left (\mbox{\boldmath $x$} 
- \mbox{\boldmath $b$}  \right )
\right ],
\label{eqn:current}
\end{equation}
where $\vec{Q}_{\alpha} \equiv e \vec{w}_{\alpha}$ is the 
abelian color-electric charge of the quark.
Here, $\mbox{\boldmath $a$}$ and $\mbox{\boldmath $b$}$ are position vectors 
of the quark and the antiquark, respectively, and $w_{\alpha}$ is the 
weight vector of SU(3) algebra,
$\vec{w}_1= \left (1/2, \sqrt{3}/6 \right ), \vec{w}_2= \left 
(-1/2, \sqrt{3}/6 \right ),
 \vec{w}_3= \left (0, -1/\sqrt{3} \right )$.
The label $\alpha =1, 2, 3$ corresponds to the color-electric charge, 
red(R), blue(B) and green(G).
According to the Gauss law, one finds the color-electric field and then 
the dual gauge field $\vec{B}_{\mu}$ is proportional to the quark 
charge $\vec{Q}_{\alpha}$\cite{Ichie1,Monden}.
For instance, when we consider the ${\rm R}$-$\bar{\rm {R}}$ system, 
the dual gauge field can be defined by using the weight vector 
as $\vec{B}_{\mu} \equiv  \vec{w}_{1} B_{\mu}^{\rm R}$.
In this system, the DGL lagrangian (\ref{eqn:DGL}) can be written as
\begin{eqnarray}
{\cal L '}_{\rm DGL} &=& - \frac{1}{3} \cdot \frac{1}{4}
\left (\partial_{\mu} B_{\nu}^{\rm R} - \partial_{\nu} B_{\mu}^{\rm R} 
\right )^2 
+ \left |\partial_{\mu} \chi_1 \right |^2 - 
\lambda \left ( \left |\chi_1 \right |^2-v^2 \right )^2 \nonumber\\
& & +
 \left | \left (\partial_{\mu}-\frac{1}{2} ig B_{\mu}^{\rm R} \right )
\chi_2 \right |^2
-\lambda \left ( \left |\chi_2 \right |^2-v^2 \right )^2\nonumber\\
& & +
 \left | \left (\partial_{\mu}+ \frac{1}{2} ig B_{\mu}^{\rm R} 
\right )\chi_3 \right |^2
-\lambda \left ( \left |\chi_3  \right |^2-v^2 \right )^2,
\label{eqn:DGL-2}
\end{eqnarray}
where we use the relation,
$\vec{\epsilon}_{\alpha} \cdot \vec{B}_{\mu} = 
\vec{\epsilon}_{\alpha} \cdot 
\vec{w}_1 B_{\mu}^{\rm R} = 
\frac{1}{2}\; \left (\;0,-1,1\; \right )\; B_{\mu}^{\rm R}$.
By considering the constraint condition of the phase of the monopole field
$\sum_{\alpha =1}^{3} arg \; \chi_{\alpha}$=0\cite{Suzuki,Suganuma},
we can write the monopole field, in this case, as
$\chi_1 = v$, $\chi_2 = {\chi^{\rm R}}^{\ast}$, $\chi_3 = \chi^{\rm R}$.
The DGL lagrangian (\ref{eqn:DGL-2}) is then given by
\begin{eqnarray}
{\cal L '}_{\rm DGL} &=& -\frac{1}{3} \cdot \frac{1}{4}
\left (\partial_{\mu} B_{\nu}^{\rm R} 
- \partial_{\nu} B_{\mu}^{\rm R} \right )^2 
\nonumber\\
& & +2 \left | \left (\partial_{\mu}+ \frac{1}{2} ig B_{\mu}^{\rm R}
\right ) \chi^{\rm R} \right |^2
-2\lambda \left ( \left |\chi^{\rm R} \right |^2-v^2 \right )^2.
\label{eqn:DGL-3}
\end{eqnarray}
With the redefinitions of the fields and the parameters,
\begin{equation}
B_{\mu}^{\rm R} \equiv \sqrt{3} B_{\mu}, 
\quad \chi^{\rm R} \equiv \chi,
\quad g \equiv \frac{2}{\sqrt{3}}\hat{g},
\quad \lambda \equiv 2 \hat{\lambda},
\quad v \equiv \frac{1}{\sqrt{2}}\hat{v},
\label{eqn:parameters}
\end{equation}
we get the final expression for the ${\rm R}$-$\bar{\rm {R}}$ system,
\begin{equation}
{\cal L '}_{\rm DGL} = -\frac{1}{4} \left (\partial_{\mu}B_{\nu}-\partial
_{\nu}B_{\mu} \right )^2+  \left | \left (\partial_{\mu}
+i\hat{g}B_{\mu} \right ) \chi \right |^2
-\hat{\lambda} \left  ( \left |\chi \right |^2-\hat{v}^2 \right )^2.
\label{eqn:DGL-4}
\end{equation}
For the other two color-singlet cases such as
the ${\rm B}$-$\bar{\rm {B}}$ and the ${\rm G}$-$\bar{\rm {G}}$ system,
one obtains the same expression 
owing to the Weyl symmetry among three color charges, ${\rm R}$,
${\rm B}$ and ${\rm G}$.
The lagrangian (\ref{eqn:DGL-4}) has the U(1) gauge symmetry
and its form coincides with the Ginzburg-Landau theory for 
superconductivity. 
This type of lagrangian has the flux-tube 
solution such as the Abrikosov vortex\cite{Nielsen}.

\par
To see this solution, we consider the field equations,
\begin{equation}
(\partial_{\mu}+i\hat{g}B_{\mu})^2 \chi =
2\hat{\lambda}\chi (\hat{v}^2 - \chi^* \chi),
\label{eqn:dgl-eq-1}
\end{equation}
\begin{equation}
\partial^{\nu} {}^{*}F_{\mu \nu} \equiv k_{\mu}
= -i \hat{g} (\chi^* \partial_{\mu} \chi 
- \chi \partial_{\mu} \chi^* )
+2 \hat{g}^2 B_{\mu}\chi^* \chi ,
\label{eqn:dgl-eq-2}
\end{equation}
\begin{equation}
{}^{*}F_{\mu \nu} \equiv \partial_{\mu} B_{\nu}- \partial_{\nu}  B_{\mu},
\end{equation}
with the proper boundary conditions that quantize the color-electric flux. 
The flux is given by
\begin{equation}
\Phi \equiv \int {}^{*}F_{\mu \nu} d \sigma^{\mu \nu} =
\oint B_{\mu}(x) dx^{\mu}, 
\label{eqn:flux}
\end{equation}
where $\sigma^{\mu \nu}$ is a two-dimensional surface element in the
Minkowski space.
By the polar decomposition of the monopole field
using two scalar variables, $\phi$ and $f$ as $\chi (x)= \phi (x) e^{i f(x)}$,
we obtain from Eq.(\ref{eqn:dgl-eq-2})
\begin{equation}
B_{\mu}= \frac{1}{2 \hat{g}^2} \frac{k_{\mu}}{\phi ^2}
-\frac{1}{\hat{g}} \partial _{\mu} f .
\label{eqn:dualgauge-current}
\end{equation}
We substitute this expression into (\ref{eqn:flux}) and 
integrate out over a large closed loop where the current $k_{\mu}$
is vanished.
Thus we get
\begin{equation}
\Phi = - \frac{1}{\hat{g}} \oint \partial_{\mu} f (x) dx^{\mu}.
\label{eqn:flux2}
\end{equation}
Since the only requirement on the phase $f(x)$ is that $\chi (x)$ should be
a single valued, the line 
integral (\ref{eqn:flux2}) does not necessarily vanish.
It means that $f(x)$ can be varied by $2\pi n$ ($n$=integer), therefore
\begin{equation}
\Phi = - \frac{2 \pi n}{\hat{g}},
\label{eqn:flux3}
\end{equation}
and the flux is quantized as a result of this condition. 
Integer $n$ is regarded as the winding number of the flux-tube 
corresponding to the topological charge.

\par
Let us consider the single flux-tube solution with translational 
invariance (it also has cylindrical symmetry) along the $z$-axis,
which is expected to appear in the $q$-$\bar{q}$ system.
In such a system, the dual gauge field and the monopole field 
can be written using the radial coordinate $r$ as
\begin{eqnarray}
\mbox{\boldmath $B$} &=& B(r)\mbox{\boldmath $e$} _\theta 
= \frac{\tilde{B}(r)}{r} \mbox{\boldmath $e$} _\theta ,
\nonumber\\
\phi &=& \phi(r),
\end{eqnarray}
and the phase is 
$f = n \theta$, where $\theta$ is the azimuth around the $z$-axis.
The differential of the phase is
$\nabla f =( n / r )\;\mbox{\boldmath $e$} _\theta$ and
its integration over a closed loop leads the flux quantization
condition (\ref{eqn:flux3}).
The color-electric field is defined by the rotation of the
dual gauge field,
\begin{equation}
\mbox{\boldmath $E$} \equiv \nabla \times \mbox{\boldmath $B$} 
=\frac{1}{r}\frac{d \tilde B(r)}{dr}\mbox{\boldmath $e$} _z 
\equiv E_z(r) \mbox{\boldmath $e$} _z ,
\end{equation}
where $\mbox{\boldmath $e$} _z$ is a unit vector along the $z$-axis.
The field equations (\ref{eqn:dgl-eq-1}) and (\ref{eqn:dgl-eq-2}) are
given by,
\begin{equation}
\frac{d^2\phi}{dr^2}+\frac{1}{r}\frac{d\phi}{dr}-\left(\frac{n-\hat{g}
\tilde{B}}{r}\right)^2\phi-2\hat{\lambda}\phi\;(\phi^2-\hat{v}^2)=0 ,
\label{eqn:feqcyl1}
\end{equation}
\begin{equation}
\frac{d^2\tilde{B}}{dr^2}-\frac{1}{r}\frac{d\tilde{B}}{dr}
+2\hat{g}\;(n-\hat{g}\tilde{B})\phi^2=0 ,
\label{eqn:feqcyl2}
\end{equation}
and the energy of the flux-tube per unit length $1/\hat{v}$ along 
the $z$-axis is obtained as
\begin{equation}
E_n=\frac{2\pi}{\hat{v}} \int_0^\infty rdr \left[\frac{1}{2}\left(\frac{1}{r}
\frac{d\tilde{B}}{dr}\right)^2+\left(\frac{d\phi}{dr}\right)^2
+\left(\frac{n-\hat{g}\tilde{B}}{r}\right)^2\phi^2+\hat{\lambda}
(\phi^2-\hat{v}^2)^2\right] .
\label{eqn:cylind-ene}
\end{equation}
Since the flux-tube solution should give a finite energy,
one finds the boundary conditions,
\begin{eqnarray}
\tilde{B}(r)= 0, \quad \phi (r) = 0 \quad {\rm as} \quad r \to 0 ,
\nonumber\\
\tilde{B}(r) = \frac{n}{\hat{g}}, \quad \phi (r) = \hat{v} \quad
{\rm as}\quad r \to \infty .
\label{eqn:boundary-cl}
\end{eqnarray}
The string tension can be defined by using the expression of the energy
(\ref{eqn:cylind-ene}) as
\begin{equation}
\sigma \equiv  \frac{E_{n=1}}{\int_0^{1/\hat{v}}dz} = E_{n=1} \hat{v}.
\end{equation}

\par
In Fig.\ref{fig:ft-single}, we show the numerical solution of the flux-tube,
the profiles of the color-electric field $E_z(r)$ and 
the monopole field $\phi (r)$ with the winding number $n$=1,
where the parameters are fixed as
\begin{equation}
\hat{g} = 2.6,                   %2.55 
\quad \hat{\lambda} = 33,        %33.3
\quad \hat{v} = 0.14\; {\rm GeV} %0.139
,
\label{dgl-parameters}
\end{equation}
or equivalently, 
$g=2.9$,                %2.94
$e=4\pi/g=4.3$,         %4.27
$\lambda=66$,           %66.6
$v=0.098 \;{\rm GeV}$   %0.0982
(see (\ref{eqn:parameters})).
These parameters reproduce 
the string tension 
$\sigma =  1.0 \; {\rm GeV/fm}$ and two characteristic mass 
scales which are presented
by $m_\chi  \equiv 2\sqrt{\hat{\lambda}}\hat{v}$ 
and $m_B \equiv  \sqrt{2}\hat{g}\hat{v}$ as
$m_{\chi}=1.6\;{\rm GeV}$ and $m_{B}=0.5\;{\rm GeV}$, respectively.
The {$m_{\chi}$ denotes the monopole mass, which is the threshold 
energy to excite the monopole in the QCD-vacuum 
corresponding to the Bogoliubov particle 
so-called ``Bogoliubon'' in the ordinary superconductor\cite{Tinkham}.
If such excitations dominate, the phase transition
is expected to occur and this value $m_{\chi}$ is regarded as 
the ultra-violet cutoff of the DGL theory.
The $m_B$ is the dual gauge mass, which determines the magnitude 
of the dual Meissner effect.
The value 0.5 GeV is supported by the recent calculation based on 
the lattice QCD using the dual formalism\cite{Atanaka}.
These inverse masses $m_\chi^{-1}$= 0.12 fm 
and $m_B^{-1}$= 0.39 fm
are regarded as the coherent length of the monopole field 
and the penetration depth of the color-electric field, respectively.
The ratio of these two lengths gives the Ginzburg-Landau (GL) parameter,
\begin{equation}
\tilde \kappa \equiv \frac{m_B^{-1}}{m_\chi^{-1}} 
= \frac{\sqrt{2 \hat{\lambda}}}{\hat{g}}.
\end{equation}
The GL-parameter plays an important role to define the vacuum 
properties, where $\tilde \kappa < 1$ describes the type-I vacuum and 
$\tilde \kappa > 1$ is the type-II vacuum.
The parameters (\ref{dgl-parameters}) lead the GL-parameter 
as $\tilde \kappa = 3.0 > 1$, which indicates that the QCD-vacuum belongs to 
the type-II vacuum\cite{Kato}. 

\par
In such a type-II vacuum, one can treat the field equations
(\ref{eqn:feqcyl1}) and (\ref{eqn:feqcyl2}) analytically
within the mean field approximation $\phi \sim \hat{v}$ with 
the cutoff $m_B = m_B \; \theta ( r - m_{\chi}^{-1})$. 
The cutoff is necessary in order to avoid the unphysical divergence 
at the core of the flux-tube.
The mean field approximation leads the dual London equation from
Eq.(\ref{eqn:feqcyl2}),
\begin{equation}
\frac{d^2\tilde{B}}{dr^2}-\frac{1}{r}\frac{d\tilde{B}}{dr}
+2\hat{g}(n-\hat{g}\tilde{B})\hat{v}^2=0,
\label{eqn:feqcyl2-2}
\end{equation}
and the replacements $r \equiv m_B^{-1} \rho$ and
$\tilde{B}(\rho) \equiv n/\hat{g} - \rho K(\rho)$
give
\begin{equation}
\frac{d^2K}{d\rho^2}+\frac{1}{\rho}\frac{dK}{d\rho}
-\left(1+\frac{1}{\rho^2}\right)K = 0 .
\end{equation}
We know this solution is described by 
the first order modified Bessel function $K_{1}(\rho)$, which 
asymptotically behaves as $K_1(\rho) \sim \sqrt{\frac{\pi}{2\rho}}e^{-\rho}$.
Thus one obtains the profiles of the dual gauge field and 
the color-electric field,
\begin{equation}
\tilde{B}(\rho) \sim \frac{n}{\hat{g}}- \rho 
\sqrt{\frac{\pi}{2\rho}}e^{-\rho}, \quad \quad
E_z(\rho) \sim \sqrt{\frac{\pi}{2 \rho}} 
e^{-\rho} .
\label{eqn:electf}
\end{equation}
The color-electric field is excluded from the vacuum and hence confined 
inside the region $\rho < 1$ ($r < m_B^{-1}$), which means the vortex-type, 
{\it i.e.} the flux-tube configuration.
Of course, these expressions are valid for the outside region of 
the cutoff $r > m_{\chi}^{-1}$.
If we want to get the whole region of the profiles with 
the arbitrary parameters, we must resort to the numerical calculations
as shown in Fig.\ref{fig:ft-single}.
In any case, the DGL theory explains the formation of the flux-tube
in the QCD-vacuum, which provides the linear confinement potential 
between the quark and the antiquark.

\par
Here we shall discuss some important features of the flux-tube.
As we can confirm, the phase of the monopole field $f = n \theta$ leads
the differential form of the flux quantization condition,
\begin{equation}
\nabla \times \nabla f = 2 \pi n \delta(x) \delta(y) 
\mbox{\boldmath $e$} _z ,
\label{eqn:singular}
\end{equation}
where the delta functions characterize the center of the flux-tube.
Thus, an essential point for the formation of the flux-tube is that 
the phase of the monopole field becomes singular at the center of 
the color-electric flux.
That is to say, if we want to obtain the flux-tube solution, 
all we have to do is to impose the singular structure on the phase.
We also find that the setting of the phase provides the boundary 
condition of the dual gauge field uniquely as is presented in 
Eq.(\ref{eqn:dualgauge-current}) that the dual gauge field 
should behave as $B_{\mu} \to -\frac{1}{\hat{g}} \partial _{\mu} f$ at
the current $k_{\mu} \simeq 0$.

%%% SECTION 3 %%%%%%%%%%%%%%%%%%%%%%%%%%%%%%%%%%%%%%%%%%%%%%%
\section{Glueball as the Flux-Tube Ring Solution}
\label{sec:ring}

\par
In this section, we consider the flux-tube ring solution that 
the ends of the flux-tube meet each other to form a circle with 
the radius $R$ as shown in Fig.\ref{fig:ring}.
The singular structure on the phase of the monopole field is 
characterized by the rotational invariance along the $z$-axis,
\begin{equation}
\nabla \times \nabla f = 2 \pi n \;\delta(r-R)\;\delta(z) 
\mbox{\boldmath $e$} _\theta ,
\label{eqn:ring-node}
\end{equation}
where $n$ is the winding number of the flux-tube composing the ring.
The fields can be written as
\begin{eqnarray}
\mbox{\boldmath $B$} &=& B_r(r,z) \mbox{\boldmath $e$} _r 
+ B_z(r,z) \mbox{\boldmath $e$} _z ,\nonumber\\
\phi &=& \phi(r,z),
\label{eqn:ring-field}
\end{eqnarray}
and the phase is determined by Eq.(\ref{eqn:ring-node}) as
$f = - n \; \tan^{-1} \left ( z/(r-R) \right )$.
The factor minus comes from the use of the cylindrical coordinate.
The field equations are obtained by substituting these 
expressions into Eqs.(\ref{eqn:dgl-eq-1}) and (\ref{eqn:dgl-eq-2}),
\begin{equation}
\frac{\partial ^2 \phi}{\partial r^2}+ \frac{\partial ^2 \phi}{\partial z^2}
+\frac{1}{r} \frac{\partial \phi}{\partial r}
- \hat{g}^2 \left ( {B'_r}^2 + {B'_z}^2 \right )\phi - 2 \hat{\lambda}
\;\phi\;(\phi^2-\hat{v}^2) = 0 ,
\end{equation}
\begin{equation}
\frac{\partial ^2 B_z}{\partial z \partial r} -
\frac{\partial ^2 B_r}{\partial z^2} + 2\hat{g}^2 B'_r \phi^2 =0 ,
\end{equation}
\begin{equation}
\frac{\partial ^2 B_r}{\partial r \partial z} -
\frac{\partial ^2 B_z}{\partial r^2} 
+\frac{1}{r} \left ( \frac{\partial B_r}{\partial z}
-\frac{\partial B_z}{\partial r} \right )
+ 2\hat{g}^2 B'_z \phi^2 =0 ,
\end{equation}
with
\begin{eqnarray}
B'_r &\equiv& B_r - \frac{\partial f }{\partial r} = B_r -
n \frac{z}{(r-R)^2+z^2}, \\
B'_z &\equiv& B_z - \frac{\partial f }{\partial z} = B_z +
n \frac{r-R}{(r-R)^2+z^2}.
\end{eqnarray}
The boundary conditions are given by
\begin{eqnarray}
&&\phi (r,z) = 0 \quad {\rm as}\quad (r,z) \to (R,0),\nonumber\\
&& B'_r (r,z)=0,\quad B'_z (r,z) =0 \quad {\rm and} \quad 
\phi (r,z)= \hat{v}
\quad {\rm as}\quad \sqrt{(r-R)^2+z^2} \to \infty.
\end{eqnarray}
For $r \to 0$, the color-electric field is required to disappear
due to the rotational symmetry around the $z$-axis.

\par
In Figs. \ref{fig:ring-el} and \ref{fig:ring-ph}, we show the 
numerical solutions of the profiles of the color-electric field and 
the monopole field as a function of the ring radius $R$.
These profiles show the tendencies of shrinking of the color-electric field 
and the monopole field as the ring radius $R$ is reduced.
Accordingly, we also obtain the effective string 
tension $\sigma_{\rm eff} (R)$ as a function of the ring radius 
as shown in Fig.\ref{fig:effective-st}.
$\sigma_{\rm eff} (R)$ is defined by
\begin{equation}
E(R) = 2 \pi R \sigma_{\rm eff} (R),
\label{eqn:ring-ene-1}
\end{equation}
where $E(R)$ is the energy of the flux-tube ring,
\begin{equation}
E(R) \!=\!  
2 \pi \!\!
\int_{0}^{\infty}\!\!\!\!\!\!rdr 
\!\!
\int_{- \infty}^{\infty}\!\!\!\!\!\!\!dz
\!
\left [ 
\frac{1}{2} \left ( \frac{\partial B_r}{\partial z} 
\!-\!
\frac{\partial B_z}{\partial r} \right )^2 
\!\!\!\! + \!\!
\left ( \frac{\partial \phi}{\partial r} \right )^2 
\!\!\!\! + \!\!
\left ( \frac{\partial \phi}{\partial z} \right )^2
\!\!\!\! +\! 
\hat{g}^2({B'_r}^2 \!\!+ \!{B'_z}^2)\phi^2
\!+\! \hat{\lambda}(\phi^2 \!- \! \hat{v}^2)^2 
\right ] .
\end{equation}
We find the string tension is effectively reduced with 
decreasing the ring radius $R$, which is considered to be caused by the 
reduction of the color-electric field.
The energy $E(R)$ decreases as the ring radius $R$ is reduced.
That is to say, the flux-tube ring solution in the DGL theory itself is
unstable and prefers to shrink, since it does not 
contain any kinetic term for the ring motion. 

\par
From the quantum mechanical point of view, such a collapse is to
be forbidden by the uncertainty principle like the hydrogen atom, where 
the kinetic term of the electron plays an important role
for the stability of the atom.
Hence, in order to get the stable ring solution for its motion,
it would be necessary to introduce the kinetic term of the ring. 
Since the flux-tube is characterized by the string-like singular structure
on the phase of the monopole field, it seems reasonable
to describe the flux-tube ring as the relativistic closed string 
with the effective string tension $\sigma_{\rm eff} (R)$ 
by using the Nambu-Goto (NG) action. 
The description is quite simple.
The NG action of the relativistic closed string with the string tension 
$\Sigma$ is written in general,
\begin{equation}
S = \int_{\tau_I}^{\tau_F} 
\!\!\!
d\tau \int_{0}^{2\pi} 
\!\!\!
d \theta
\; \left [ -\Sigma \; \sqrt{(\dot{X} X')^2-(\dot{X})^2 ( X')^2} \;\right ],
\end{equation}
where $X^{\mu} = X^{\mu}( \tau ,\theta )$ denotes the string world sheet,
$\dot{X}^{\mu} \equiv \partial X^{\mu} / \partial \tau$ and
${X^{\mu}}' \equiv \partial X^{\mu} / \partial \theta$.

\par
We parameterize the ring as a circle with the radius $R$,  
\begin{equation}
X^1(\tau,\theta)=R(\tau) \cos \theta ,\quad
X^2(\tau,\theta)=R(\tau) \sin \theta ,
\label{eqn:parameterization}
\end{equation}
and choose the chronological gauge $X^{0}(\tau,\theta) \equiv \tau $. 
This parameterization satisfies the orthogonal 
condition $\dot{ \mbox{\boldmath $X$} } \cdot \mbox{\boldmath $X$}' = 0$.
Thus, we obtain the action of the flux-tube ring,
\begin{equation}
S_{\rm ring} = \int_{\tau_I}^{\tau_F}\!\!\! 
d\tau \int_{0}^{2\pi} \!\!\! d \theta
\; \left [ - \sigma_{\rm eff} (R) R \sqrt{1- \dot{R}^2} \;\right ],
\end{equation}
and the hamiltonian of the ring, 
\begin{equation}
H(P_R, R) = \sqrt{P_R^2+ \{ 2 \pi R \sigma_{\rm eff} (R) \}^2 } ,
\label{eqn:mass}
\end{equation}
where $P_R$ is the canonical conjugate momentum of the coordinate $R$,
defined by
\begin{equation}
P_R \equiv 2\pi R \sigma_{\rm eff} (R) \frac{\dot{R}}{\sqrt{1-\dot{R}^2}}.
\end{equation}
If we put $P_R = 0$ ($\dot{R}=0$), the hamiltonian provides
the static energy (\ref{eqn:ring-ene-1}).

\par
Once the ring hamiltonian including the kinetic term is obtained, we 
can look for the glueball states by solving 
the Schr$\ddot{\rm o}$dinger equation
\begin{equation}
\left [ - \frac{d^2}{dR^2} + \{ 2 \pi R \sigma_{\rm eff} (R) \}^2 \right ]
 \;\Phi_m (R) = {M_m}^2 \; \Phi_m (R),
\label{eqn:wave-eq}
\end{equation}
with the boundary conditions,
\begin{equation}
\Phi_m (R=0)=0, \quad \Phi_m (R=\infty)=0.
\end{equation}
The boundary condition $\Phi_m (0)$=0 is required in terms of the ring 
structure of the flux-tube since the wave function is considered to 
characterize the configuration of the color-electric flux.

\par
It is useful to consider the type-II limit where the effective 
string tension has a constant value; $\sigma_{\rm eff} (R) 
\approx \sigma$ ($\simeq$ 1.0 GeV/fm).
In this case, the ring hamiltonian reduces into the harmonic-oscillator 
in one dimension and we can easily obtain the analytic form of 
the wave function and the mass spectrum,
\begin{eqnarray}
&&\Phi_m (R) \propto H_m (\sqrt{2 \pi \sigma} R ) 
\exp ( - \pi \sigma R^2 ),\\
&&M_m = \sqrt{4 \pi \sigma \left(m + \frac{1}{2} \right)},
\end{eqnarray}
where $H_m(x)$ is Hermite polynomials, $H_{0}(x)=1$, $H_{1}(x)=x$ and
so on.
One finds the state $m=1,3,5,\cdots$ satisfy the boundary 
condition $\Phi_m (0)$=0.
Thus, we get
\begin{eqnarray}
&&\Phi_1 (R) = 2^{7/2} \pi \sigma^{3/2}
R  \exp ( - \pi \sigma R^2 ),\\
&&M_1 = \sqrt{6 \pi \sigma} = 4.34 \sqrt{\sigma} = 1.93 \;{\rm GeV},
\label{eqn:analytic-spectrum}
\end{eqnarray}
for the lowest state of the flux-tube ring. 
The root mean square radius is obtained as
\begin{equation}
\sqrt{\langle {R_1}^{2} \rangle} 
\equiv \int_0^{\infty} dR \Phi_1 R^2 \Phi_1
= \sqrt{\frac{3}{4 \pi \sigma}} 
= 0.489 \frac{1}{\sqrt{\sigma}}
= 0.23 \;{\rm fm}.
\label{eqn:analytic-radius}
\end{equation}

\par
Let us calculate the ground state of the $m=1$ state for 
the $\tilde \kappa =3.0$ case.
In this case, we should resort to the variational method 
since the effective string tension is not a constant value and is 
given as the numerical function of the radius $R$.
We use the trial function
$\Phi_1(R, a) \propto R  \exp ( - a \pi \sigma R^2 )$
 where $a$ is the variational parameter determined by minimizing
\begin{equation}
M_1(a) \equiv \sqrt{\frac{\langle \Phi_1(R, a) | H(P_R, R)^2 | 
\Phi_1(R, a) \rangle}{\langle \Phi_1(R, a) | 
\Phi_1(R, a) \rangle }}  ,
\end{equation}
and we obtain $M_1(a = 0.82 )= 1.6 \;{\rm GeV}$
as shown in Fig.\ref{fig:glueball-mass},
which is regarded as the lowest glueball mass $M_G$.
As for the root mean square radius, $a = 0.82 < 1$ suggests that 
the ring radius becomes broad compared with $\sqrt{3/4 \pi \sigma}$ for 
the type-II limit case by the factor $1/\sqrt{a}$. 
Therefore, we estimate the ring radius as $0.25\;{\rm fm}$ and the size 
of the glueball as $R_G = 0.25 \times 2 = 0.5\;{\rm fm}$ (the ring diameter).
We find that this mass spectrum $M_G$=1.6 GeV is almost consistent with 
the scalar glueball mass that the lattice QCD predicts for the 
lowest state\cite{Bali,Chen,Sexton,Teper,Morningstar}.

\par
It is interesting to note that the expression (\ref{eqn:analytic-spectrum})
is very similar to the following form\cite{Burakovski},
\begin{equation}
M(0^{++}) = 3 \sqrt{2} \sqrt{\sigma} \simeq 4.24 \sqrt{\sigma},
\label{eqn:naive-scalar}
\end{equation}
which is naively derived by the procedure of the minimization 
of the energy of a bound state of two massless gluons,
\begin{equation}
E = 2 p + \frac{9}{4} \sigma r - \frac{\alpha}{r},
\end{equation}
where $p$ is the gluon momentum and $\alpha$ the strong coupling
constant.
The color factor 9/4 is given by the ratio of the 
SU($N_c$) Casimir operators of 
the adjoint representation $N_c$ and the fundamental 
representation $({N_c}^2-1)/2 N_c$ for $N_c$=3.
The uncertainty relation $p \cdot r \geq 1$ leads the energy minimum
$E=3\sqrt{(2-\alpha)\sigma} \approx 3\sqrt{2} \sqrt{\sigma}$
at $r=2\sqrt{2-\alpha}/ 3\sqrt{\sigma} \approx 
2\sqrt{2}/ 3\sqrt{\sigma} = 0.943 / \sqrt{\sigma}$.
One may find that this glueball size 0.943/$\sqrt{\sigma}$ is also consistent 
with two times of 0.489/$\sqrt{\sigma}$ in (\ref{eqn:analytic-radius}).
These similarities seem to suggest a close relation
between the flux-tube ring picture and the phenomenological potential 
picture of the glueball.

%%% SECTION 4 %%%%%%%%%%%%%%%%%%%%%%%%%%%%%%%%%%%%%%%%%%%%%%%
\section{Summary and discussions}
\label{sec:summary}

\par
We have studied the flux-tube ring solution in the dual 
Ginzburg-Landau (DGL) theory as the glueball excitation.
The flux-tube solution in the DGL theory explains the color confinement
and also provides intuitive pictures of the hadrons in terms of the 
string-like structure of the color-electric flux.
The hadrons including the valence quarks topologically 
correspond to the open flux-tube excitation with terminals.
Thus, the glueball, which is considered as an object without valence 
quarks, can be regarded as the flux-tube ring intuitively.

\par
By considering the rotational invariant system along the $z$-axis
as shown in Fig.\ref{fig:ring}, we have studied the profiles of the 
color-electric field and the monopole field as a function of the ring radius.
We have used the parameters which reproduce $m_B$=0.5 GeV, 
 $m_\chi$=1.6 GeV and the string tension $\sigma$=1.0 GeV/fm.
The GL-parameter is found to be $\tilde \kappa$=3.0, which suggests that 
the QCD-vacuum belongs to the type-II vacuum.
We have calculated the effective string
tension $\sigma_{\rm eff} (R)$ as a function of the ring radius.
$\sigma_{\rm eff} (R)$ is defined by the relation
$E(R)\! = \! 2 \pi R \sigma_{\rm eff} (R)$,
where $E(R)$ is the energy of the ring with the radius $R$.
We have found the profiles are reduced with decreasing the ring radius $R$
and accordingly the effective string tension is reduced.
These results characterize the size effect of the flux-tube, which is 
the difference between the flux-tube and the string.

\par
In order to include the kinetic term of the ring,
we have described the flux-tube ring as the relativistic closed string
with the effective string tension. Using the Nambu-Goto (NG) action,
we have parameterized the ring as a circle with the radius $R$ and
obtained the hamiltonian
$H(P_R, R) \! =  \! \sqrt{P_R^2+\{2\pi R \sigma_{\rm eff} (R) \}^2}$,
where $P_R$ is the canonical conjugate momentum of the coordinate $R$.
If we put $P_R=0$, the hamiltonian leads the static energy $E(R)$.
Analyzing the Schr$\ddot{\rm o}$dinger equation
$H(P_{R},R)^{2} \Phi (R)\!\! = \!\! {M}^2 \Phi (R)$
with the boundary condition $\Phi (R=0)\!\! =\!\! \Phi (R=\infty)\!\! =\!\! 0$,
we have obtained the eigenvalue $M_G \! =\!\! 1.6\;{\rm GeV}$ for the 
ground state, which is considered as the lowest glueball mass.
The size of the glueball is estimated as $R_G \! =\!\! 0.5\;{\rm fm}$.
The mass spectrum $M_G \! =\!\! 1.6\;{\rm GeV}$ is 
almost consistent with the scalar glueball mass that the lattice QCD 
predicts for the lowest state.
We have found these results are very similar to another
approach based on the Regge phenomenology, where
the color factor 9/4 in the linear potential between two gluons plays 
important roles for the estimation of the glueball mass and the size.
These similarities are quite interesting and
the phenomenological potential picture of the glueball 
%which is including the color factor 
seems to have a close relation with the flux-tube ring picture.

\par
Here, we shall discuss about the relation between the glueball 
and the monopole.
One may find that the $m_\chi$=1.6 GeV is very similar to the 
glueball mass that we have obtained above analysis.
The monopole field denotes a complex scalar field
and its origin is the off-diagonal gluon field in the MA-gauge in QCD.
Thus, the monopole field would also present the scalar gluonic excitation
in the QCD-vacuum such as the scalar glueball\cite{Suzuki}.
Therefore, this resemblance of masses seems to be quite natural, in fact,
the phase of the monopole field has played an essential role for the 
flux-tube ring solution.
It is interesting to note that once this identification is allowed, we 
can determine the mass $m_\chi$ self-consistently. 
In such a case, the DGL theory which is now including three-parameters 
can be rewritten to the two-parameters theory.
However, whether the both scalar glueballs presented by the 
flux-tube ring or the monopole field are the same or not is another 
problem since the flux-tube ring depends not only on the GL-parameter 
but also on the string tension.
We are now investigating the scalar glueball in terms of the monopole field
in the DGL theory.

\par
Again, we would like to mention that our main idea is the description 
of the flux-tube ring solution 
in the DGL theory as the relativistic closed string with the effective 
string tension, which enables us to write the hamiltonian of the 
flux-tube ring using the NG action. Once the hamiltonian is obtained, we 
can discuss the mass spectrum and the wave function of the glueball state.
The boundary condition $\Phi (R=0)$=0 dictates the ring structure of the 
color-electric flux to the wave function.
In the future, we should consider the collective motion of the 
ring and extract the physical glueball state 
with definite quantum numbers $J^{PC}$ using the angular 
momentum projection method.
Although such approaches are in progress, we can expect that the DGL 
theory provides a useful method for the study of the glueball.

%%% SECTION 5 %%%%%%%%%%%%%%%%%%%%%%%%%%%%%%%%%%%%%%%%%%%%%%%
\section*{Acknowledgment}

\par
We would like to thank all of the members of the RCNP theory group 
for useful comments and discussions.
One of the authors (H.S.) is supported in part by Grant for
Scientific Research (No.09640359) from the Ministry of Education,
Science and Culture, Japan.

%%% SECTION 6 %%%%%%%%%%%%%%%%%%%%%%%%%%%%%%%%%%%%%%%%%%%%%%%
%\section*{References}

\def\Journal#1#2#3#4{{#1} {\bf #2}, #3 (#4)}
%For Phys. Rev format.
\def\NCA{Nuovo Cimento}
\def\NPB{Nucl. Phys. {\bf B}}
\def\PLB{Phys. Lett. B}
\def\PRL{Phys. Rev. Lett.}
\def\PRD{Phys. Rev. D}
\def\PRC{Phys. Rev. C}
\def\ZPC{Z. Phys. C}
\def\PTP{Prog. Theor. Phys.}
\def\PTPS{Prog. Theor. Phys. (Suppl.)}
\def\JP{J.Phys. G:Nucl. Part. Phys.}
\def\SJ{Sov. J. of Nucl. Phys.}
\def\NPBPS{Nucl. Phys. B (Proc. Suppl.)}

%%% SECTION 7 %%%%%%%%%%%%%%%%%%%%%%%%%%%%%%%%%%%%%%%%%%%%%%%
\newpage
\section*{Figure Captions}

\par
FIG.1 :
Profiles of the color-electric field $E_z(r)$ (dotted) and 
the monopole field $\phi (r)$ (solid) of the cylindrical flux-tube 
in the type-II ($\tilde \kappa = 3.0$) vacuum as functions of the radial
distance from the center of the flux-tube $r$, where
the parameters are fixed as
$\hat{g} = 2.6,%2.55
\hat{\lambda} = 33,%33.3
\hat{v} = 0.14\; {\rm GeV}%0.139
$.

\vspace{0.4cm}
\par
FIG.2 :
The flux-tube ring system which has rotational invariance 
along the $z$-axis. $R$ denotes the ring radius.
All the coordinates used in the text are defined in this figure.

\vspace{0.4cm}
\par
FIG.3 :
The profiles of the color-electric field 
$E_{\theta}(r,z)$ in unit of $1/{\rm fm^2}$ of the flux-tube ring system
in the type-II ($\tilde \kappa = 3.0$) vacuum.
The left-hand side denotes the 3D plot and the right-hand side
is its contour plot.
The unit of the radial coordinate $r$ and the $z$-axis is fm.
The radius is taken from 2.0 fm (upper) to 0.5 fm (below) in
step of 0.5fm.
The color-electric field $E_{\theta}$ decreases as the ring 
radius $R$ is reduced.

\vspace{0.4cm}
\par
FIG.4 :
The profiles of the monopole field $\phi(r,z)$ in unit of $1/{\rm fm}$ 
of the flux-tube ring system
in the type-II ($\tilde \kappa = 3.0$) vacuum.
The left-hand side denotes the 3D plot and the right-hand side
is its contour plot.
The unit of the radial coordinate $r$ and the $z$-axis is fm.
The radius is taken from 2.0 fm (upper) to 0.5 fm (below) 
in step of 0.5fm.
The monopole field  $\phi$ at the central region of the ring decreases
as the ring radius $R$ is reduced.

\vspace{0.4cm}
\par
FIG.5 :
Effective string tension $\sigma_{\rm eff} (R)$ in GeV/fm as a 
function of the ring radius $R$.
As the ring radius is reduced, the effective string tension 
decreases to zero.

\vspace{0.4cm}
\par
FIG.6 :
The energy expectation value $M_1 (a)$ of the flux-tube ring system
as a function of the variational parameter $a$.
The dotted line denotes the case of the constant string tension
$\sigma =1.0\;{\rm GeV/fm}$ (for type-II limit), where
the energy minimum shows 1.93 GeV at $a =1$ as we have 
obtained in the analytical way.
The solid line is the main result by using the effective string tension 
$\sigma_{\rm eff} (R)$ (for $\tilde \kappa = 3.0$), which 
shows the energy minimum 1.60 GeV at $a =0.82$.
The result $a < 1$ suggests that the wave function is broad compared with
the type-II limit.

%%% Figures %%%%%%%%%%%%%%%%%%%%%%%%%%%%%%%%%%%%%%%%%%%%%%%%
%% Section.2 %%
\newpage
\begin{figure}[hbt]
%{\Large Fig.1}\\
\begin{center}
\begin{minipage}[t]{12.0cm}
\epsfxsize=12.0cm
\vspace{4cm}
\epsfbox{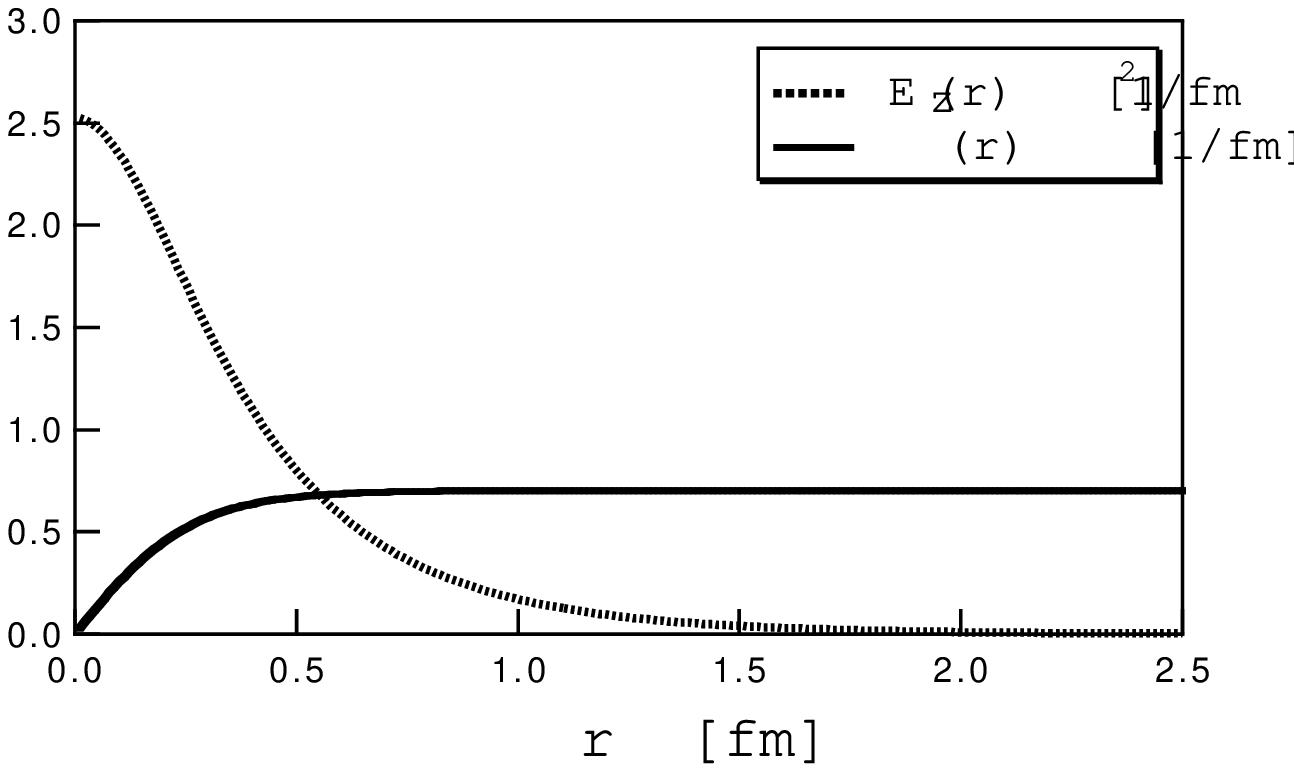}
\caption[Cylindrical flux-tube solution in the DGL theory]{}
\label{fig:ft-single}
\end{minipage}
\end{center}
\end{figure}

%% Section.3 %%
\newpage
\begin{figure}[hbt]
%{\Large Fig.2}\\
\begin{center}
\begin{minipage}[t]{14.0cm}
\epsfxsize=14.0cm
\epsfysize=6.9cm
\vspace{4cm}
\epsfbox{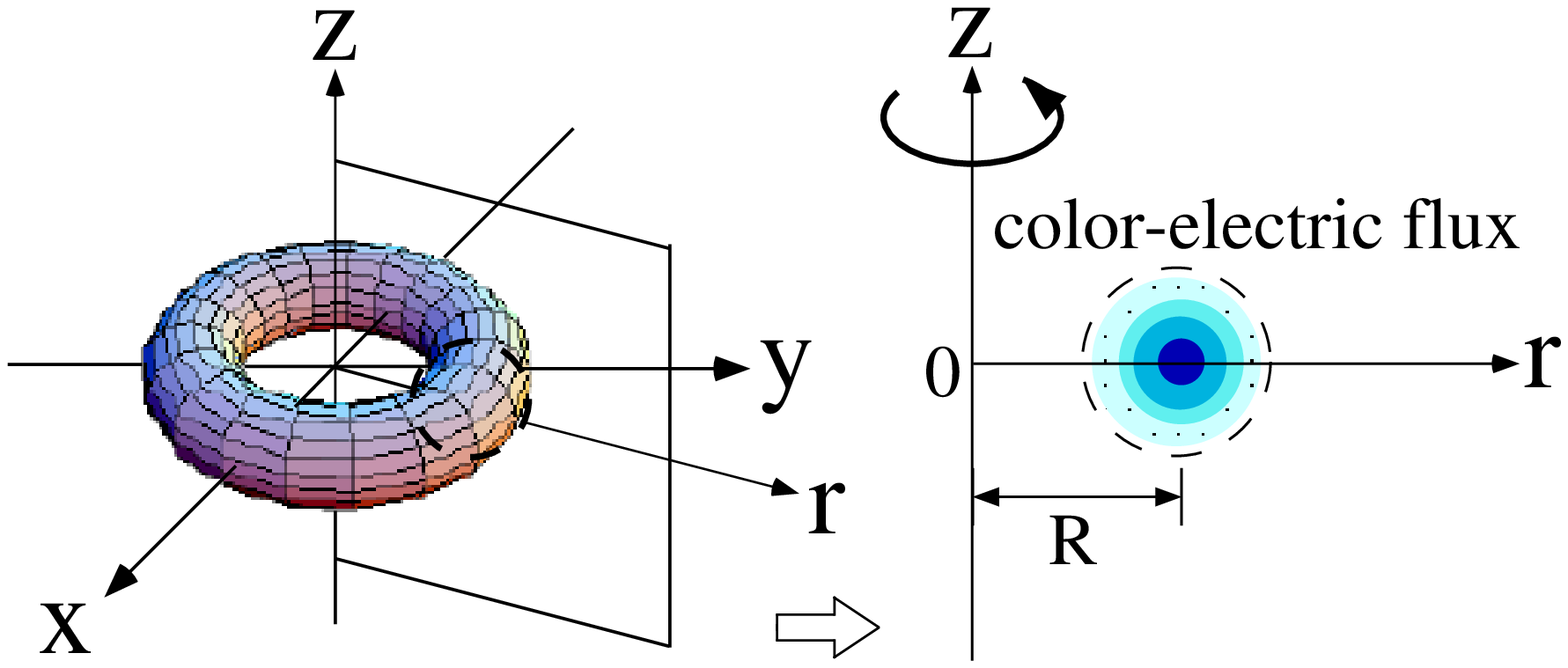}
\caption[Flux-tube ring]{}
\label{fig:ring}
\end{minipage}
\end{center}
\end{figure}

\newpage
\begin{figure}[t]
%{\Large Fig.3}\\
\begin{center}
\begin{minipage}[t]{12.56cm}
\epsfxsize=13cm
\epsfysize=21.45cm
\epsfbox{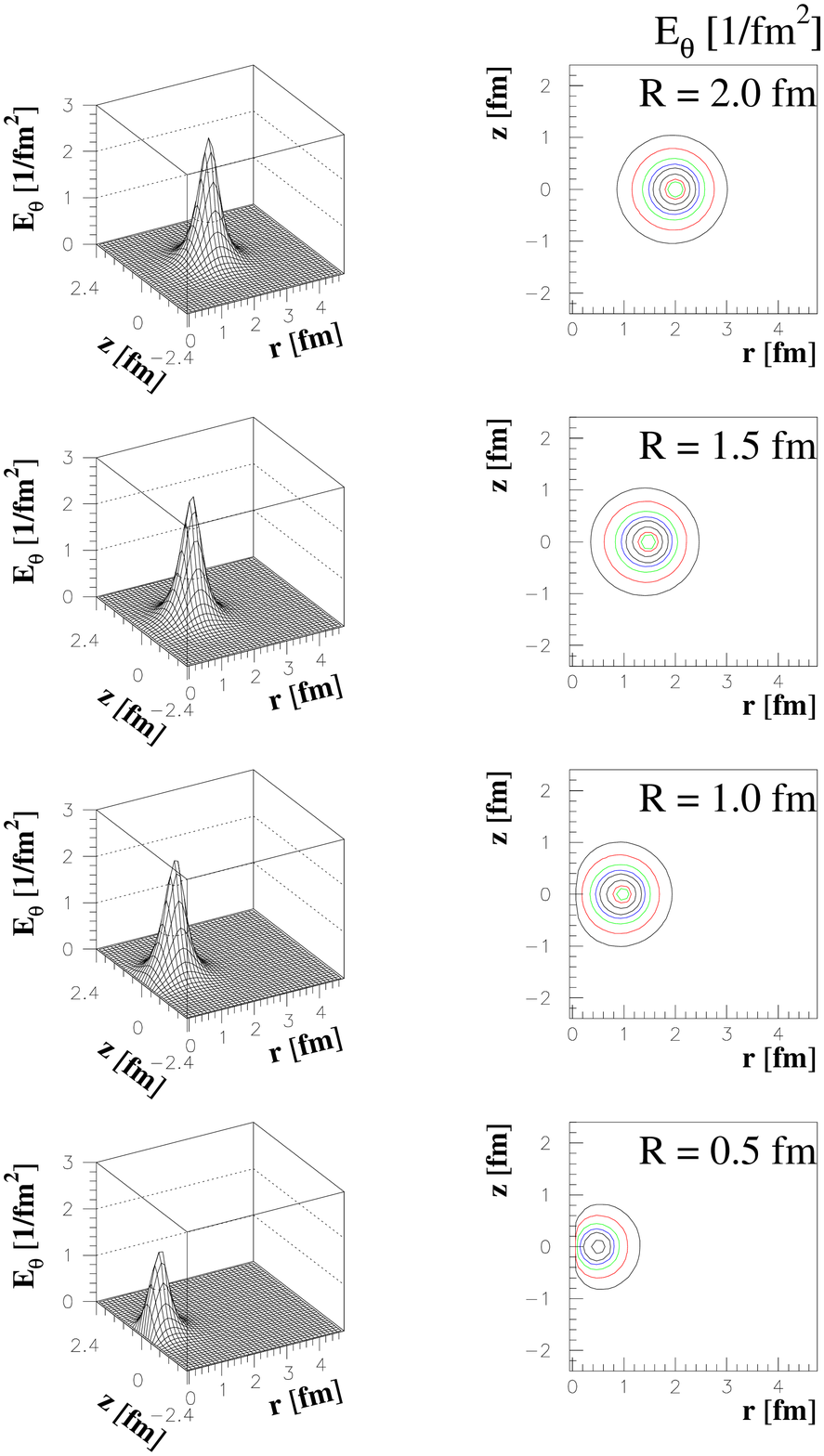}
\end{minipage}
\begin{minipage}[t]{12.56cm}
\caption[Profiles of the color-electric field in the flux-tube ring]{}
\label{fig:ring-el}
\end{minipage}
\end{center}
\end{figure}

\newpage
\begin{figure}[t]
%{\Large Fig.4}\\
\begin{center}
\begin{minipage}[t]{12.56cm}
\epsfxsize=13cm
\epsfysize=21.45cm
\epsfbox{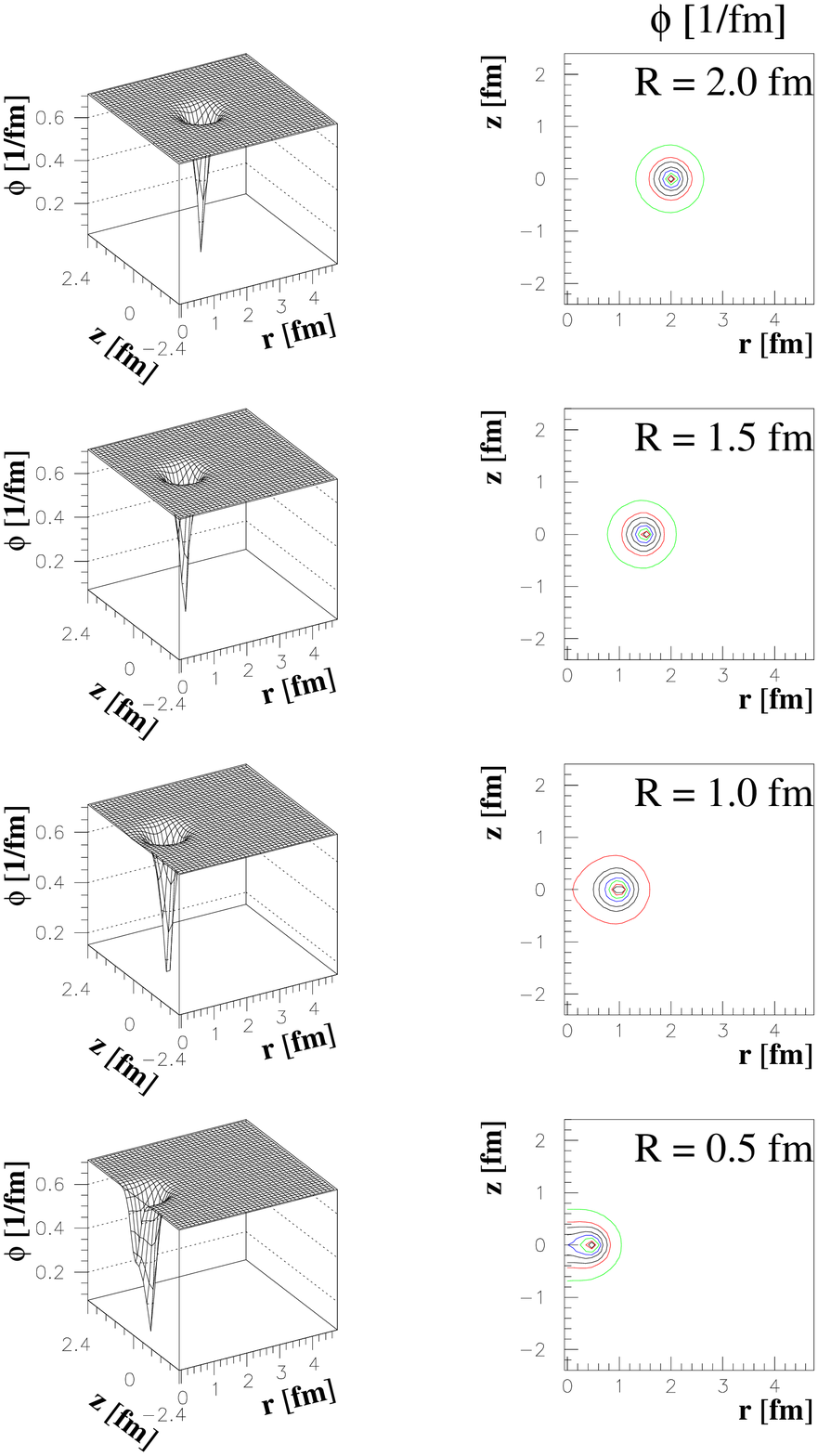}
\end{minipage}
\begin{minipage}[t]{12.56cm}
\caption[Profiles of the monopole field in the flux-tube ring]{}
\label{fig:ring-ph}
\end{minipage}
\end{center}
\end{figure}

\newpage
\begin{figure}[hbt]
%{\Large Fig.5}\\
\begin{center}
\begin{minipage}[t]{12.0cm}
\epsfxsize=12.0cm
\vspace{4cm}
\epsfbox{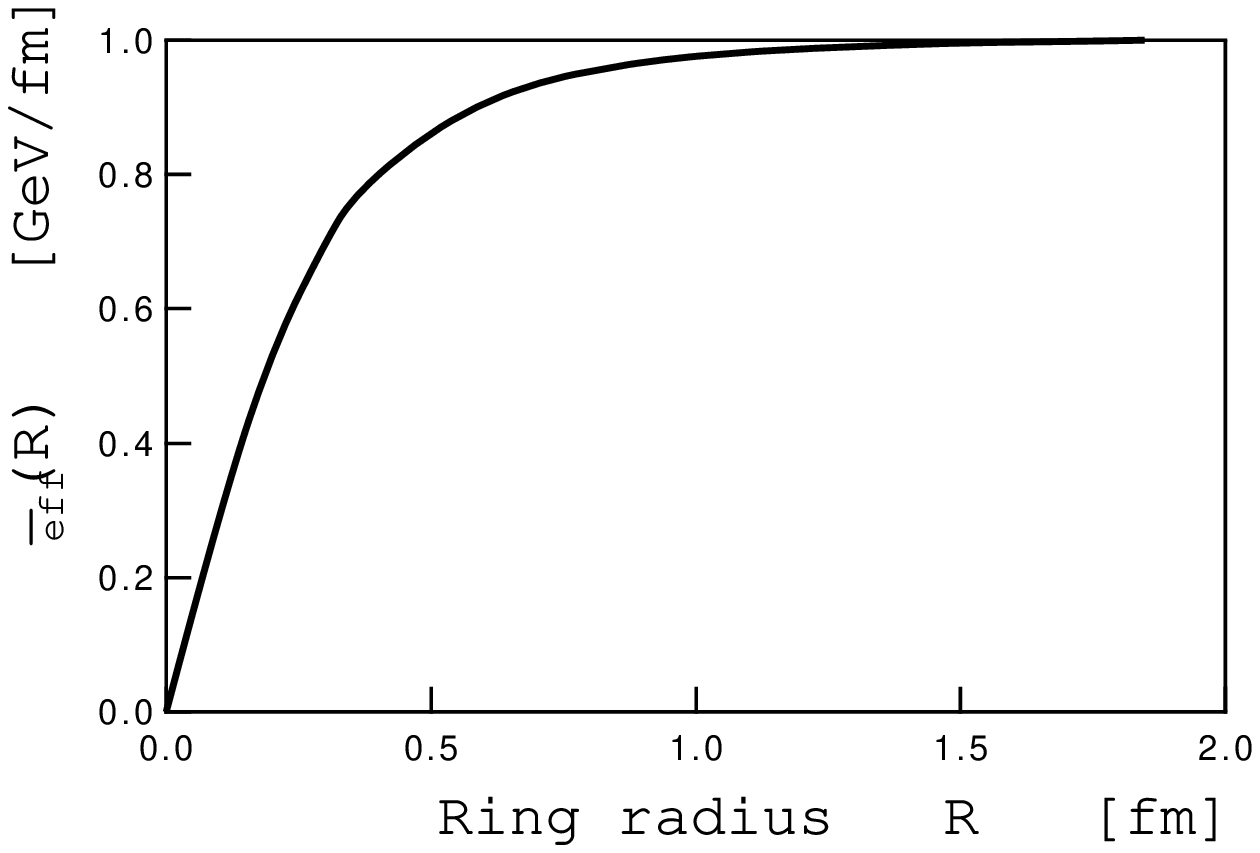}
\caption[String tension v.s ring radius]{}
\label{fig:effective-st}
\end{minipage}
\end{center}
\end{figure}

\newpage
\begin{figure}[hbt]
%{\Large Fig.6}\\
\begin{center}
\begin{minipage}[t]{12cm}
\epsfxsize=12cm
\vspace{4cm}
\epsfbox{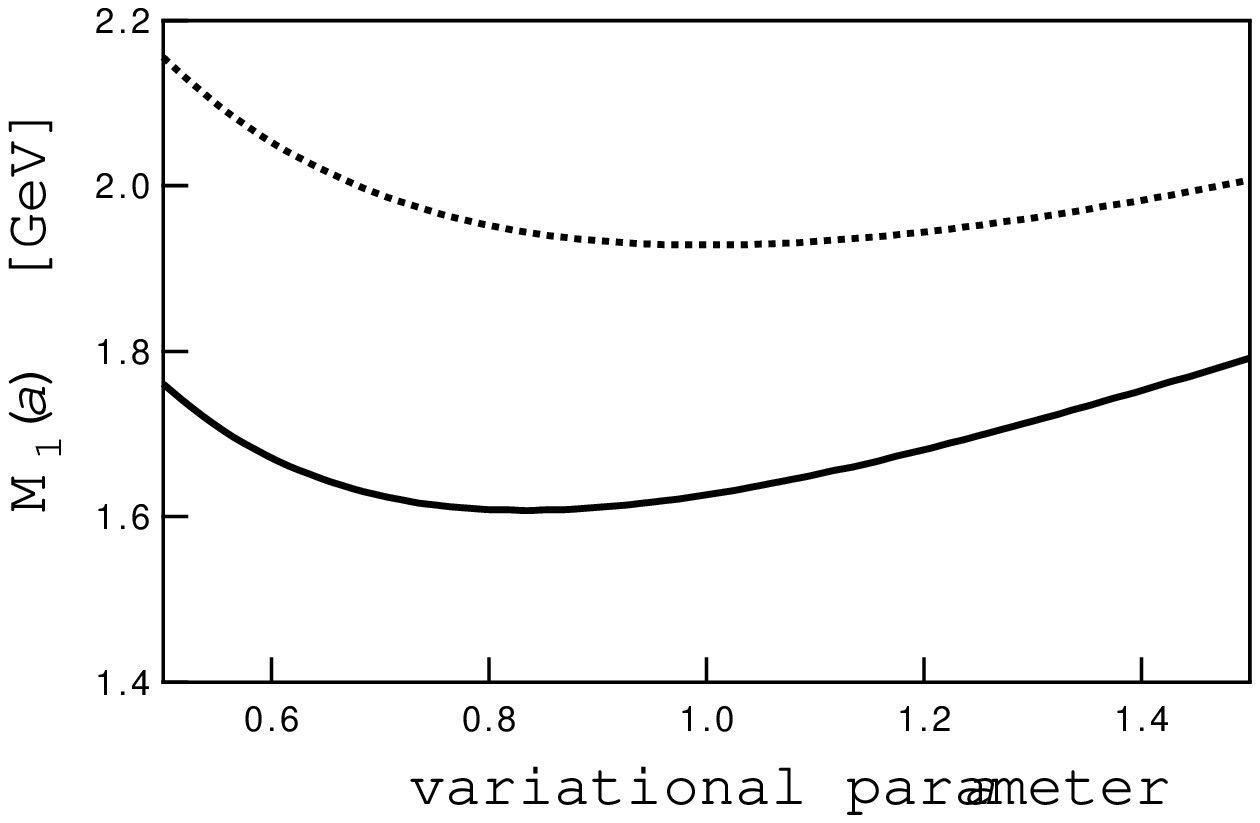}
\caption[Glueball Mass]{}
\label{fig:glueball-mass}
\end{minipage}
\end{center}
\end{figure}

\end{document}